\documentclass{article}
\usepackage{amssymb}
\usepackage{amsmath}
\usepackage{amsfonts}
\usepackage{amsthm}

\newtheorem{theorem}{Theorem}[section]
\newtheorem{lemma}[theorem]{Lemma}
\newtheorem{corollary}[theorem]{Corrolary}
\newtheorem{remark}[theorem]{Remark}
\input{tcilatex}
\begin{document}

\title{On the connection between the solutions to\ the Dirac and Weyl
equations and the corresponding electromagnetic 4-potentials}
\author{Aristides I. Kechriniotis \thanks{%
Department of Physics, University of Thessaly, GR-35100 Lamia, Greece.
Corresponding author. E-mail: arisk7@gmail.com } \and Christos A. Tsonos 
\thanks{%
Department of Physics, University of Thessaly, GR-35100 Lamia, Greece } \and %
Konstantinos K. Delibasis \thanks{%
Department of Computer Science and Biomedical Informatics, University of
Thessaly, GR-35100 Lamia, Greece } \and Georgios N. Tsigaridas\thanks{%
Department of Physics, School of Applied Mathematical and Physical Sciences,
National Technical University of Athens,GR-15780 Zografou Athens, Greece }}
\date{May 22, 2022}
\maketitle

\begin{abstract}
In this work we study in detail the connection between the solutions to the
Dirac and Weyl equations and the associated electromagnetic 4-potentials.
First, it is proven that all solutions to the Weyl equation are degenerate,
in the sense that they correspond to an infinite number of electromagnetic
4-potentials. As far as the solutions to the Dirac equation are concerned,
it is shown that they can be classified into two classes. The elements of
the first class correspond to one and only one 4-potential, and are called
non-degenerate Dirac solutions. On the other hand, the elements of the
second class correspond to an infinite number of 4-potentials, and are
called degenerate Dirac solutions. Further, it is proven that at least two
of these 4-potentials are gauge-inequivalent, corresponding to different
electromagnetic fields. In order to illustrate this particularly important
result we have studied the degenerate solutions to the force-free Dirac
equation and shown that they correspond to massless particles. We have also
provided explicit examples regarding solutions to the force-free Weyl
equation and the Weyl equation for a constant magnetic field. In all cases
we have calculated the infinite number of different electromagnetic fields
corresponding to these solutions. Finally, we have discussed potential
applications of our results in cosmology, materials science and
nanoelectronics.

\end{abstract}


\section{Introduction}

\numberwithin{equation}{section}%
The Dirac equation has been the first electron equation in quantum mechanics
to satisfy the Lorentz covariance \cite{Dirac}, \cite{Hestenes}, initiating
the beginning of one of the most powerful theories ever formulated: the
quantum electrodynamics. This equation predicted the spin and the magnetic
moment of the electrons, the existence of antiparticles and was able to
reproduce accurately the spectrum of the hydrogen atom. The Dirac equation
plays an important role in various fields of Physics, as mathematical
physics \cite{Kerner}, \cite{Breev}, \cite{Oertel}, particle physics \cite%
{Sogut}, \cite{Okninski},\cite{Saeedi}, solid state physics \cite{Rakhimov}, 
\cite{Fukuyama-Fuseya-Ogata-Kobayashi-Suzumura}, astrophysics \cite{Batic},
quantum computing \cite{Fillion-Gourdeau}, non-linear optics \cite%
{Qin-Feng-Liu}, etc. As it can be deduced from the cited articles the Dirac
equation and its applications is still a hot topic of research.

\bigskip The majority of the previously reported works focus on the
determination of the wave function $\Psi $\ when the electromagnetic
4-potential is given. However, the inverse problem, as formulated by Eliezer 
\cite{Eliezer} is also quite interesting: \textquotedblleft Given the wave
function $\Psi $, what can we say about the electromagnetic potential $%
A_{\mu }$, which is connected to $\Psi $\ by Dirac's equation? Is $A_{\mu }$%
\ uniquely determined, and if not, what is the extent to which it is
arbitrary?\textquotedblright . In his relevant work Eliezer found an
expression for the magnetic vector $A_{\mu }$ and the electric scalar
potential $\phi $ as a function of $\Psi $. In another important work by
Radford et al \ \cite{Radford}, the Dirac equation is expressed in a
2-spinor form, which allows it to be (covariantly) solved for the
electromagnetic 4-potential, in terms of the wave function and its
derivatives. This approach subsequently led to some physically interesting
results \cite{Booth-Radford}, \cite{Radford-Booth}, \cite{Booth 1} (for a
review see \cite{Booth 2}). Further, in \cite{Booth-Legg-Jarvis} it was
demonstrated that the Dirac equation is indeed algebraically invertible if a
real solution for the 4-potential is required. Namely, two expressions for
the components $A_{\mu }$ of the electromagnetic 4-potential are presented,
equivalent to the one given in \cite{Radford}. However, these expressions
are not valid in the case that $\Psi ^{\dagger }\gamma ^{0}\Psi ,$ $\Psi
^{\dagger }\gamma ^{1}\gamma ^{2}\gamma ^{3}\Psi $ are identically zero,
where $\gamma ^{\mu }$, $\mu =0,1,2,3$, is the stsandard representation of
the Dirac matrices, which are explicitely provided in the following section.
Thus, in the above mentioned works,the existence of nonzero Dirac solutions
satisfying the aforementioned conditions is not studied. Further, if it is
assumed that these solutions exist, it should be investigated if one such
solution corresponds to a unique real 4-potential or not. Consequently, the
inversion of the Dirac equation is an open problem, which will be fully
solved in the present article.

The main results of this paper are the following. First, it is proven that
all solutions to the Weyl equations are degenerate, in the sense that they
correspond to an infinite number of electromagnetic 4-potentials, which are
explicitely calculated. This result is particularly important from a
practical point of view as it is expected to offer new possibilities
regarding the technological applications of certain materials as graphene
sheets, Weyl semimetals, etc., where ordinary charged particles can
collectively behave as massless \cite{Novoselov}, \cite{Ciudad}, \cite{Park}%
, \cite{Kamal}, \cite{Lai}, \cite{Armitage}.

Further, it is thoroughly proven that every Dirac spinor corresponds to one
and only one mass.This result is utilized to show that the set of all Dirac
spinors can be classified into two classes. The elements of the first class
correspond to one and only one 4-potential, and are called non-degenerate
Dirac solutions, while the elements of the second class correspond to an
infinite number of 4-potentials, and are called degenerate Dirac solutions.
In the first case the 4-potential is fully defined as a function of the
Dirac spinor, while in the second one explicit expressions are provided for
the infinite number of 4-potentials corresponding to the degenerate Dirac
spinors. Further, it is proven that at least two of these 4-potentials are
gauge-inequivalent, corresponding to different electromagnetic fields. From
a physical point of view this is quite a surprising result, meaning that
particles described by a specific class of spinors can be in the same state
under the influence of different electromagnetic fields.

In order to illustrate this particularly important result we study the
degeneracy condition for the force-free Dirac equation and found that it is
satisfied for massless particles. Thus, a massless Dirac particle can exist
in the same state both in the absence of any electromagnetic field as well
as under the influence of an infinite number of different electromagnetic
fields, which are explicitely calculated. This remarkable result could be
used to explain the rapid evolution of the Universe during the firts stages
of its creation, before the particles acquire mass \cite{Gromov}, \cite%
{Gorbunov}. We also show that the degeneracy can be extended to massive
particles if particle-antiparticle pairs are considered. Further, we provide
explicit examples of Weyl spinors, and calculate the infinite number of
different electromagnetic fields corresponding to these solutions. Finally,
we show that the state of massless Dirac and Weyl particles does not change
if a static or time-dependent electric field is applied parallel (or
anti-parallel) to their direction of motion. This important result implies
that Ohm's law does not hold for massless Dirac and Weyl particles, and the
current they transfer remains constant, unaffected by the applied electric
field.

\section{Preliminaries}

\bigskip The Dirac equation for a fermion of charge $q$ and mass $m$
described by the spinor wavefunction $\Psi $ in the presence of an external
electromagnetic 4-potential $A_{\mu },$ can be written as \cite{Radford}%
\begin{equation}
a_{\mu }\gamma ^{\mu }\Psi =-\left( i\gamma ^{\mu }\partial _{\mu }\Psi
-m\Psi \right) ,  \tag{2.1}
\end{equation}%
\ where \ $a_{\mu }=qA_{\mu }$, $\gamma ^{0}=\left[ 
\begin{array}{cc}
\sigma ^{0} & 0 \\ 
0 & -\sigma ^{0}%
\end{array}%
\right] $ and $\gamma ^{\mu }=\left[ 
\begin{array}{cc}
0 & \sigma ^{\mu } \\ 
-\sigma ^{\mu } & 0%
\end{array}%
\right] $, $\mu =1,2,3.$ Here it is assumed that we are working in the
natural system of units where $\hbar =c=1.$We have also used the notation $%
\left( \sigma ^{0},\sigma ^{1},\sigma ^{2},\sigma ^{3}\right) =\left(
I_{2},\sigma _{x},\sigma _{y},\sigma _{z}\right) $, where $I_{2}$ is the
two-dimensionnal identity matrix and $\sigma _{x}=\left[ 
\begin{array}{cc}
0 & 1 \\ 
1 & 0%
\end{array}%
\right] ,\sigma _{y}=\left[ 
\begin{array}{cc}
0 & -i \\ 
i & 0%
\end{array}%
\right] ,\sigma _{z}=\left[ 
\begin{array}{cc}
1 & 0 \\ 
0 & -1%
\end{array}%
\right] $ are the well-known Pauli matrices.

In the following we shall also use the two Weyl equations under the
electromagnetic potential $A_{\mu },$ which can be written in the form \cite%
{Stepanovskiy},%
\begin{equation}
a_{\mu }\sigma ^{\mu }\psi =-i\sigma ^{\mu }\partial _{\mu }\psi ,  \tag{2.2}
\end{equation}%
and $2a_{0}\sigma ^{0}\psi -a_{\mu }\sigma ^{\mu }\psi =-\left( 2i\sigma
^{0}\partial _{0}\psi -i\sigma ^{\mu }\partial _{\mu }\psi \right) .$

describing massless particles with their spin parallel to their propagation
direction (positive helicity) and anti-parallel to their propagation
direction (negative helicity) respectively. In the rest of the article, we
shall also assume that the electromagnetic 4-potential is always real.

\begin{definition}
Any solution of equation (2.1) for a 4-potential $a_{\mu }$ and a mass $m$
will be called Dirac solution , and any solution of equation $(2.2)$ or $%
(2.3)$ for a 4-potential $a_{\mu }$ will be called Weyl solution .
\end{definition}

\begin{definition}
A Dirac solution $\Psi $ is said to correspond to a mass $m$ , if there
exists a 4-potential $a_{\mu }$, such that $\Psi $ is a solution of $(2.1)$
for this mass $m.$
\end{definition}

\begin{definition}
A Dirac solution $\Psi $ is said to correspond to a 4-potential $a_{\mu }$,
if there exists a mass $m$, such that $\Psi $ is a solution of $(2.1)$ for
this 4-potential $a_{\mu }.$
\end{definition}

\begin{definition}
A Dirac solution $\Psi $ is said to correspond to an elecromagnetic field $%
\left( \mathbf{E},\mathbf{B}\right) $, if $\ \Psi $ corresponds to a
4-potential $a_{\mu }$ associated to $\left( \mathbf{E},\mathbf{B}\right) .$
\end{definition}

\begin{definition}
A Weyl solution $\psi $ is said to correspond to a potential $a_{\mu }$, if $%
\psi $ is a solution of $(2.2)$~or $\left( 2.3\right) $ for this 4-potential 
$a_{\mu }.$
\end{definition}

\begin{itemize}
\item We denote the following matrices 
\begin{equation*}
\gamma ^{5}:=i\gamma ^{0}\gamma ^{1}\gamma ^{2}\gamma ^{3},\text{\ \ \ }%
\gamma :=\gamma ^{0}+\gamma ^{0}\gamma ^{5}
\end{equation*}
\end{itemize}

\begin{remark}
Since $\gamma ^{0}$ is Hermitian and $\gamma ^{0}\gamma ^{5}$ is
anti-Hermitian, we have that $\Psi ^{\dagger }\gamma ^{0}\Psi $ is real and $%
\Psi ^{\dagger }\gamma ^{0}\gamma ^{5}\Psi $ is imaginary. Therefore the
equation $\Psi ^{\dagger }\gamma \Psi =0$ is equivalent to $\ \Psi ^{\dagger
}\gamma ^{0}\Psi =\Psi ^{\dagger }\gamma ^{0}\gamma ^{5}\Psi =0$ .
\end{remark}

\bigskip In \cite{Booth-Legg-Jarvis} two equivalent algebraic expressions
for the 4-potential in terms of the Dirac solution are derived:

\begin{equation}
\alpha _{\mu }=\frac{1}{2}\frac{i\left( \overline{\Psi }\gamma ^{\mu }\NEG%
{\partial}\Psi -\overline{\Psi }\overleftarrow{\NEG{\partial}}\gamma ^{\mu
}\Psi \right) -2mj_{\mu }}{\overline{\Psi }\Psi },  \tag{2.4}
\end{equation}

\begin{equation}
\alpha _{\mu }=\frac{i}{2}\frac{\overline{\Psi }\gamma ^{5}\gamma ^{\mu }\NEG%
{\partial}\Psi -\overline{\Psi }\gamma ^{5}\overleftarrow{\NEG{\partial}}%
\gamma ^{\mu }\Psi }{\overline{\Psi }\gamma ^{5}\Psi }  \tag{2.5}
\end{equation}%
where $\overline{\Psi }:=\Psi ^{\dagger }\gamma ^{0}$, $j_{\mu }:=$ $%
\overline{\Psi }\gamma ^{\mu }\Psi $, and $\overline{\Psi }\gamma ^{\nu }%
\overleftarrow{\NEG{\partial}}\gamma ^{\mu }\Psi :=\overline{\NEG%
{\partial}\Psi }\gamma ^{\nu }\gamma ^{\mu }\Psi .$

\begin{description}
\item[Notation] \textit{Let }$f,g$\textit{\ be any functions. Then }$f\equiv
g$\textit{\ \ means that }$f$\textit{\ is identically equal to }$g.$
\end{description}

\begin{remark}
Clearly, from equations $\left( 2.4\right) $, $\left( 2.5\right) $ and the
remark 2.1 follows that if $\Psi ^{\dagger }\gamma \Psi \not\equiv 0$ , then 
$\Psi $ correspods to one and only one real 4-potential $\alpha _{\mu }$. In
more detail if $\overline{\Psi }\Psi \not\equiv 0$ and $\overline{\Psi }%
\gamma ^{5}\Psi \equiv 0$, then the potential is given by $\left( 2.4\right) 
$. On the other hand, if $\overline{\Psi }\Psi \equiv 0$ and $\overline{\Psi 
}\gamma ^{5}\Psi \not\equiv 0$, then the potential is given by $\left(
2.5\right) $.~Finally, if $\overline{\Psi }\Psi \not\equiv 0$ and $\overline{%
\Psi }\gamma ^{5}\Psi \not\equiv 0$, then according to \cite%
{Booth-Legg-Jarvis} both formulas $\left( 2.4\right) $ and $\left(
2.5\right) $ are equivalent and so the potential is given either by $\left(
2.4\right) $ or $\left( 2.5\right) $.
\end{remark}

However, these inversion formulas are valid only in the case where $%
\overline{\Psi }\Psi \not\equiv 0$ or $\overline{\Psi }\gamma ^{5}\Psi
\not\equiv 0$. Therefore, a question arises here, concerning the existence
of Dirac solutions $\Psi \not\equiv 0$\ satisfying the conditions $\overline{%
\Psi }\Psi \equiv \overline{\Psi }\gamma ^{5}\Psi \equiv 0$, or equivalently 
$\Psi ^{\dagger }\gamma \Psi \equiv 0$. First, do such solutions exist, and
if yes, do they correspond to one and only one real 4-potential or not? The
main aim of the rest of this article is to answer this question. The above
considerations lead us to introduce the following definition:

\begin{definition}
A Dirac or Weyl solution $\Psi $ is said to be degenerate, if and only if $\
\Psi $corresponds to more than one 4-potentials.
\end{definition}

\section{On the degeneracy of Weyl spinors}

In this section we shall prove that all solutions to the Weyl equations in
the form (2.2) and (2.3) are degenerate, in the sense that they correspond
to an infinite number of electromagnetic 4-potentials.

\begin{theorem}
Any non identically zero Weyl spinor is degenerate,\ corresponding to an
infinite number of electromagnetic 4-potentials $b_{\mu }$ given by the
formulae 
\begin{equation}
\left( b_{0}.b_{1}.b_{2},b_{3}\right) =\left( a_{0}+f\phi _{0},a_{1}+f\phi
_{1},a_{2}+f\phi _{2},a_{3}+f\phi _{3}\right) ,  \tag{3.1}
\end{equation}%
and 
\begin{equation}
\left( b_{0}.b_{1}.b_{2},b_{3}\right) =\left( a_{0}+f\theta
_{0},a_{1}+f\theta _{1},a_{2}+f\theta _{2},a_{3}+f\theta _{3}\right) , 
\tag{3.2}
\end{equation}%
for equations (2.2) and (2.3) respectively. Here $f$ is any real function of 
$\ $the variables $x_{\mu }$, and $\phi _{\mu }$, $\theta _{\mu }$ are real
functions given by%
\begin{equation*}
\left( \phi _{0},\phi _{1},\phi _{2},\phi _{3}\right) =\left( 1,-\frac{\psi
^{\dagger }\sigma ^{1}\psi }{\psi ^{\dagger }\psi },-\frac{\psi ^{\dagger
}\sigma ^{2}\psi }{\psi ^{\dagger }\psi },-\frac{\psi ^{\dagger }\sigma
^{3}\psi }{\psi ^{\dagger }\psi }\right) ,
\end{equation*}%
and 
\begin{equation*}
\left( \theta _{0},\theta _{1},\theta _{2},\theta _{3}\right) =\left( 1,%
\frac{\psi ^{\dagger }\sigma ^{1}\psi }{\psi ^{\dagger }\psi },\frac{\psi
^{\dagger }\sigma ^{2}\psi }{\psi ^{\dagger }\psi },\frac{\psi ^{\dagger
}\sigma ^{3}\psi }{\psi ^{\dagger }\psi }\right) .
\end{equation*}
\end{theorem}

\begin{proof}
First we will show that any solution $\psi $ of the first Weyl equation $%
\left( 2.2\right) $ corresponds to any potential $b_{\mu }$ as given by $%
\left( 3.1\right) :$ Clearly, equation $\left( 2.2\right) $ can be written
as 
\begin{equation}
(a_{\mu }+f\phi _{\mu })\sigma ^{\mu }\psi =-i\partial _{\mu }\sigma ^{\mu
}\psi +f\phi _{\mu }\sigma ^{\mu }\psi   \tag{3.3}
\end{equation}%
It is easy to verfy that for any two spinor $\psi $ the following identity
holds:%
\begin{equation}
\phi _{\mu }\sigma ^{\mu }\psi =0.  \tag{3.4}
\end{equation}%
Then, from $\left( 3.3\right) $ and $\left( 3.4\right) $ we conclude that $%
\psi $ corresponds to $b_{\mu }$.

Next we will try to extract all 4-potentials $b_{\mu }$ corresponding to $%
\psi $ from the equation $\left( 2.2\right) $ :

Since $\psi $ corresponds also to $b_{\mu }$ we have 
\begin{equation}
\sigma ^{\mu }b_{\mu }\psi =-i\sigma ^{\mu }\partial _{\mu }\psi  \tag{3.5}
\end{equation}%
Substracting $\left( 2.2\right) $ from $\left( 3.5\right) $ we obtain 
\begin{equation}
\sigma ^{\mu }\left( b_{\mu }-a_{\mu }\right) \psi =0.  \tag{3.6}
\end{equation}%
Multiplying $\left( 3.6\right) $ from the left succesivelly by $\psi ^{\dag
}\sigma ^{1}$, $\psi ^{\dag }\sigma ^{2}$, $\psi ^{\dag }\sigma ^{3}$ ,
adding the resulting three equations with their hermitian conjugates, and
taking into account that the matrices $\sigma ^{\mu }$ are hermitian and $%
\sigma ^{1}\sigma ^{2}$, $\sigma ^{2}\sigma ^{3}$, $\sigma ^{3}\sigma ^{1}$
are antihermitian, \ we obtain the following set of equations: 
\begin{eqnarray*}
\left( b_{0}-a_{0}\right) \psi ^{\dag }\sigma ^{1}\psi +\left(
b_{1}-a_{1}\right) \psi ^{\dag }\psi &=&0, \\
\left( b_{0}-a_{0}\right) \psi ^{\dag }\sigma ^{2}\psi +\left(
b_{2}-a_{2}\right) \psi ^{\dag }\psi &=&0, \\
\left( b_{0}-a_{0}\right) \psi ^{\dag }\sigma ^{3}\psi +\left(
b_{3}-a_{3}\right) \psi ^{\dag }\psi &=&0,
\end{eqnarray*}%
Finally, setting $b_{0}=a_{0}+f,$ we get $(3.1)$. Further, from the
hermiticity of $\sigma ^{\mu }$ follows that the functions $\phi _{1}$,$%
~\phi _{2}$, $\phi _{3}$ are real. Consequently, all 4 -potentials given by $%
\left( 3.1\right) $ are real.

Working exactly as above we can prove \ that any solution $\psi $ of $\left(
2.3\right) $ is also solution to the second Weyl equation under any of the
4-potentials $b_{\mu }$ given by $\left( 3.2\right) $, which are also real.
\end{proof}

This is a particularly interesting result not only from a theoretical, but
also from a practical point of view, because it has been shown recently \cite%
{Novoselov}, \cite{Ciudad}, \cite{Park}, \cite{Kamal}, \cite{Lai}, \cite%
{Armitage}, that ordinary charged particles can also behave as massless in
certain materials as graphene sheets, Weyl semimetals, etc. Therefore, the
property of Weyl particles to be in the same state under a wide variety of
different electromagnetic fields is expected to be offer new possibilities
regarding the practical applications of these materials, which are already
important \cite{Garcia}, \cite{Gray}, \cite{Hills}. In section 7 we provide
explicit examples of Weyl spinors, calculating also the electromagnetic
fields corresponding to these solutions.

\section{Uniqueness of mass}

\bigskip In this section we shall prove that any Dirac spinor corresponds to
one and only one value of the mass. This result is essential to prove the
main theorem of this paper regarding degenerate Dirac solutions, which is
presented in the next section. In \cite{Eliezer} and \cite{Booth-Legg-Jarvis}
there was given the following formula for the mass $m$ in terms of spinor $%
\Psi $:

\begin{equation*}
m=-\frac{i}{2}\frac{\overline{\Psi }\gamma ^{5}\overleftarrow{\NEG{\partial}}%
\Psi +\overline{\Psi }\gamma ^{5}\NEG{\partial}\Psi }{\overline{\Psi }\gamma
^{5}\Psi },
\end{equation*}%
which clearly is not valid in the case where $\Psi ^{\dagger }\gamma
^{1}\gamma ^{2}\gamma ^{3}\Psi =0$.

In this section we prove that any not identically zero Dirac solution
corresponds to one and only one mass. This result will also be used in the
next section.

\begin{definition}
In the set of all Dirac solutions we define the following relation:
\end{definition}

\begin{itemize}
\item $\Psi _{1}\approx \Psi _{2}$ if and only if $\;\Psi _{1}$,$\Psi _{2}$
are gauge equivalent, that is there exists a non zero number $c$ and a
differentiable function of the spatial and temporal variables $f:\mathbb{R}%
^{4}\rightarrow \mathbb{R}$ such that $\Psi _{1}=ce^{if}\Psi _{2}\text{.}$
Clearly $""\approx ""$ is an equivalence relation, and by $\left[ \Psi %
\right] $ we will denote the equivalence class of $\Psi \text{.}$

\item We denote by $B\left( \Psi \right) $ the set of all masses to which $%
\Psi $ corresponds.
\end{itemize}

\begin{lemma}
Let $\Psi \not\equiv 0$ be a Dirac solution, then for any $\Psi _{1}\in %
\left[ \Psi \right] $ we have $B\left( \Psi _{1}\right) =B\left( \Psi
\right) .$
\end{lemma}

\begin{proof}
From $\Psi _{1}$ $\in $ $\left[ \Psi \right] $ we have $\Psi =c_{1}\exp
\left( if\right) \Psi _{1}$, for some not identically zero $c_{1}\in \mathbb{%
R}$ and some differentiable function $f:\mathbb{R}^{4}\rightarrow \mathbb{R}%
\text{.}$ So from (2.1) we get

\begin{equation*}
i\gamma ^{\mu }\partial _{\mu }\Psi _{1}-m\Psi _{1}=\left( a_{\mu }-\partial
_{\mu }f\right) \gamma ^{\mu }\Psi 
\end{equation*}%
Therefore, $\Psi _{1}$ corresponds to the mass $m$ and to the 4-potential 
\begin{equation*}
b_{\mu }=a_{\mu }-\partial _{\mu }f\text{.}
\end{equation*}
\end{proof}

\begin{corollary}
\textit{Any not identically zero Dirac solution }$\Psi $ \textit{corresponds
to one and only one mass.}
\end{corollary}

\begin{proof}
Let $\Psi $ be a Dirac solution corresponding to a 4-potential $\alpha _{\mu
}$ and a mass $m$. We choose $\Psi _{1}\in \left[ \Psi \right] $ defined by $%
\Psi =\exp \left( if\right) \Psi _{1}$, where 
\begin{equation}
f:=\int_{k}^{x_{0}}a_{0}\left( s,x_{1},x_{2},x_{3}\right) ds  \tag{4.1}
\end{equation}

where  $k$ is a real constant. Then according to Lemma 4.1, $\Psi _{1}$
corresponds to the potential $b_{\mu }=a_{\mu }-\partial _{\mu }f$ and the
mass $m$, and from $\left( 4.1\right) $ we have $b_{0}=a_{0}-\partial
_{0}f=a_{0}-\alpha _{0}=0.$ Now we apply the formulas $\left( 2.4\right) $
for the spinor $\Psi _{1}$ by writting only the first component of $b_{\mu }$%
, and using the fact that that $b_{0}=0$: 
\begin{equation*}
m=\frac{i}{2}\frac{\overline{\Psi _{1}}\gamma ^{0}\NEG{\partial}\Psi _{1}-%
\overline{\Psi _{1}}\overleftarrow{\NEG{\partial}}\gamma ^{0}\Psi _{1}}{\Psi
_{1}^{\dagger }\Psi _{1}}.
\end{equation*}%
which, since $\Psi _{1}\not\equiv 0$, means that the above formula is well
defined. Therefore $\Psi _{1}$ corresponds to a unique mass~$m$ which,
according to Lemma 4.1, means that $\Psi $ also corresponds to a unique mass~%
$m$.
\end{proof}

\section{On degenerate Dirac solutions}

In this section we shall prove the main theorem of this paper, according to
which all Dirac solutions are classified into two classes. Degenerate
solutions corresponding to an infinite number of 4-potentials and
non-degenerate solutions corresponding to one and only one 4-potential. The
4-potentials are explicitly calculated in both cases. Further, as far as the
degenerate solutions are concerned, it is proven that at least two
4-potentials are gauge inequivalent, and consequently correspond to
different electromagnetic fields. In order to proceed with these results,
the following Lemmas are required:

\begin{lemma}
Let $\Psi $ be any spinor. Then we have 
\begin{align}
\left( \Psi ^{T}\gamma ^{2}\Psi \right) \overline{\left( \Psi ^{T}\gamma
^{0}\gamma ^{1}\gamma ^{2}\Psi \right) }& =\left( \Psi ^{T}\gamma ^{0}\gamma
^{1}\gamma ^{2}\Psi \right) \overline{\left( \Psi ^{T}\gamma ^{2}\Psi
\right) }  \tag{5.1} \\
\left( \Psi ^{T}\gamma ^{2}\Psi \right) \overline{\left( \Psi ^{T}\gamma
^{0}\Psi \right) }& =\left( \Psi ^{T}\gamma ^{0}\Psi \right) \overline{%
\left( \Psi ^{T}\gamma ^{2}\Psi \right) }  \notag \\
\left( \Psi ^{T}\gamma ^{2}\Psi \right) \overline{\left( \Psi ^{T}\gamma
^{0}\gamma ^{2}\gamma ^{3}\Psi \right) }& =\left( \Psi ^{T}\gamma ^{0}\gamma
^{2}\gamma ^{3}\Psi \Psi \right) \overline{\left( \Psi ^{T}\gamma ^{2}\Psi
\right) }  \notag
\end{align}%
if and only if 
\begin{equation}
\left( \Psi ^{T}\gamma ^{2}\Psi \right) \left( \Psi ^{\dagger }\gamma \Psi
\right) =0\text{.}  \tag{5.2}
\end{equation}
\end{lemma}

\begin{proof}
The above lemma will be proved pointwise. We set 
\begin{equation}
\Psi =\left[ 
\begin{array}{cccc}
1 & 1 & 0 & 0 \\ 
0 & 0 & 1 & 1 \\ 
1 & -1 & 0 & 0 \\ 
0 & 0 & 1 & -1%
\end{array}%
\right] \left[ 
\begin{array}{c}
\zeta _{1} \\ 
\zeta _{2} \\ 
\zeta _{3} \\ 
\zeta _{4}%
\end{array}%
\right] \Leftrightarrow \left[ 
\begin{array}{c}
\zeta _{1} \\ 
\zeta _{2} \\ 
\zeta _{3} \\ 
\zeta _{4}%
\end{array}%
\right] =\frac{1}{2}\left[ 
\begin{array}{cccc}
1 & 0 & 1 & 0 \\ 
1 & 0 & -1 & 0 \\ 
0 & 1 & 0 & 1 \\ 
0 & 1 & 0 & -1%
\end{array}%
\right] \Psi ,  \tag{5.3}
\end{equation}%
where $\Psi =\left[ 
\begin{array}{cccc}
\psi _{1}\left( x\right) & \psi _{2}\left( x\right) & \psi _{3}\left(
x\right) & \psi _{4}\left( x\right)%
\end{array}%
\right] ^{T},$ and $x$ is any element in the domain of $\Psi $ $.$ After
some algebra, the relations (5.1) can be written as 
\begin{align*}
\left( \zeta _{1}\zeta _{4}-\zeta _{2}\zeta _{3}\right) \overline{\left(
\zeta _{1}\zeta _{2}-\zeta _{3}\zeta _{4}\right) }& = & & \overline{\left(
\zeta _{1}\zeta _{4}-\zeta _{2}\zeta _{3}\right) }\left( \zeta _{1}\zeta
_{2}-\zeta _{3}\zeta _{4}\right) \text{,} \\
\left( \zeta _{1}\zeta _{4}-\zeta _{2}\zeta _{3}\right) \overline{\left(
\zeta _{1}\zeta _{2}+\zeta _{3}\zeta _{4}\right) }& = & & \overline{\left(
\zeta _{1}\zeta _{4}-\zeta _{2}\zeta _{3}\right) }\left( \zeta _{1}\zeta
_{2}+\zeta _{3}\zeta _{4}\right) \text{,} \\
\zeta _{1}\zeta _{4}\overline{\zeta _{2}\zeta _{3}}& = & & \overline{\zeta
_{1}\zeta _{4}}\zeta _{2}\zeta _{3}\text{,}
\end{align*}%
which by setting 
\begin{equation}
\zeta _{1}=\omega _{1}\zeta _{3}\text{ and }\zeta _{4}=\omega _{2}\zeta _{2}
\tag{5.4}
\end{equation}%
take the following form: 
\begin{align*}
\left\vert \zeta _{2}\zeta _{3}\right\vert ^{2}\left[ \left( \omega
_{1}\omega _{2}-1\right) \left( \overline{\omega _{1}}-\overline{\omega _{2}}%
\right) -\left( \overline{\omega _{1}\omega _{2}}-1\right) \left( \omega
_{1}-\omega _{2}\right) \right] & = & & 0\text{,} \\
\left\vert \zeta _{2}\zeta _{3}\right\vert ^{2}\left[ \left( \omega
_{1}\omega _{2}-1\right) \left( \overline{\omega _{1}}+\overline{\omega _{2}}%
\right) +\left( \overline{\omega _{1}\omega _{2}}-1\right) \left( \omega
_{1}+\omega _{2}\right) \right] & = & & 0\text{,} \\
\left\vert \zeta _{2}\zeta _{3}\right\vert ^{2}\left( \omega _{1}\omega _{2}-%
\overline{\omega _{1}\omega _{2}}\right) & = & & 0.
\end{align*}%
Consequently 
\begin{equation*}
\zeta _{2}=0\text{ or }\zeta _{3}=0\;\text{or }\left\{ 
\begin{array}{c}
\left( \omega _{1}\omega _{2}-1\right) \left( \overline{\omega _{1}}-%
\overline{\omega _{2}}\right) -\left( \overline{\omega _{1}\omega _{2}}%
-1\right) \left( \omega _{1}-\omega _{2}\right) =0\text{,} \\ 
\left( \omega _{1}\omega _{2}-1\right) \left( \overline{\omega _{1}}+%
\overline{\omega _{2}}\right) +\left( \overline{\omega _{1}\omega _{2}}%
-1\right) \left( \omega _{1}+\omega _{2}\right) =0\text{,} \\ 
\omega _{1}\omega _{2}\in \mathbb{R}\text{,}%
\end{array}%
\right\}
\end{equation*}%
which is equivalent to 
\begin{equation*}
\zeta _{2}=0\text{ or }\zeta _{3}=0\;\text{or }\omega _{1}\omega _{2}-1=0%
\text{ or }\left\{ 
\begin{array}{c}
\overline{\omega _{1}}-\overline{\omega _{2}}-\omega _{1}+\omega _{2}=0\text{%
,} \\ 
\overline{\omega _{1}}+\overline{\omega _{2}}+\omega _{1}+\omega _{2}=0\text{%
,}%
\end{array}%
\right\}
\end{equation*}%
or 
\begin{equation*}
\zeta _{2}=0\text{ or }\zeta _{3}=0\text{ or }\omega _{1}\omega _{2}-1=0%
\text{ or }\omega _{1}+\overline{\omega _{2}}=0\text{,}
\end{equation*}%
or 
\begin{equation*}
\left\vert \zeta _{2}\right\vert ^{2}\zeta _{3}^{2}\left( \omega _{1}\omega
_{2}-1\right) \left( \omega _{1}+\overline{\omega _{2}}\right) =0\text{,}
\end{equation*}%
Using (5.4) the above relations take the form 
\begin{equation*}
\left( \zeta _{1}\zeta _{4}-\zeta _{2}\zeta _{3}\right) \left( \zeta _{1}%
\overline{\zeta _{2}}+\zeta _{3}\overline{\zeta _{4}}\right) =0\text{,}
\end{equation*}%
which through (5.3) can be rewritten as (5.2).
\end{proof}

\begin{lemma}
Let $\Psi $ be any spinor. Then we have,%
\begin{equation}
\Psi ^{T}\gamma ^{2}\Psi =\Psi ^{\dagger }\gamma \Psi =0\text{,}  \tag{5.5}
\end{equation}%
if and only if \ 
\begin{equation*}
\Psi =\left[ 
\begin{array}{c}
\psi \\ 
\psi%
\end{array}%
\right] \;\text{or }\Psi =\text{ }\left[ 
\begin{array}{c}
\psi \\ 
-\psi%
\end{array}%
\right] \text{,}
\end{equation*}%
where $\psi $ is a two spinor.
\end{lemma}

\begin{proof}
The present lemma will also be proved pointwise. For any element \ $x$ in
the domain of $\Psi $, setting (5.3) in (5.5) we get 
\begin{equation}
\zeta _{1}\zeta _{4}=\zeta _{2}\zeta _{3}\text{, }\overline{\zeta _{1}}\zeta
_{2}+\overline{\zeta _{3}}\zeta _{4}=0\text{,}  \tag{5.6}
\end{equation}%
We suppose that for some $x$ in the domain of $\Psi $, holds that $\zeta
_{1}\zeta _{2}\zeta _{3}\zeta _{4}\neq 0$. Then, from (5.6) we easily obtain 
\begin{equation*}
\QOVERD\vert \vert {\zeta _{1}}{\zeta _{3}}^{2}=-1.
\end{equation*}%
Therefore for any $x$ in the domain of $\Psi $ holds that $\zeta _{1}\zeta
_{2}\zeta _{3}\zeta _{4}=0$. Hence, from (5.6) we have that for any $x$ in
the domain of $\Psi $ 
\begin{equation*}
\zeta _{1}=\zeta _{3}=0\text{ or }\zeta _{2}=\zeta _{4}=0,
\end{equation*}%
which using (5.3) can be rewritten as 
\begin{equation*}
\psi _{4}+\psi _{2}=\psi _{3}+\psi _{1}=0\text{ or }\psi _{4}-\psi _{2}=\psi
_{3}-\psi _{1}=0.
\end{equation*}
\end{proof}

\begin{lemma}
Let $\Psi $ $\not\equiv 0$ be a Dirac solution correpondig to a mass $m$ and
to a 4-potential $a_{\mu }$. Then we have $\Psi ^{T}\gamma ^{2}\Psi \equiv
\Psi ^{\dagger }\gamma \Psi \equiv 0$ if and only if $\ $either $\Psi =\left[
\begin{array}{c}
\psi \\ 
\psi%
\end{array}%
\right] $, where $\psi $ is a solution to the Weyl equation in the form $%
\left( 2.2\right) $ or $\Psi =\left[ 
\begin{array}{c}
\psi \\ 
-\psi%
\end{array}%
\right] $, where $\psi $ is a solution to the Weyl equation in the form $%
\left( 2.3\right) $. correponding to the 4-potential $a_{\mu }$.
\end{lemma}

\begin{proof}
Setting $\Psi =\left[ 
\begin{array}{c}
\psi \\ 
\psi%
\end{array}%
\right] $ in $\left( 2.1\right) $ we equivalently get the following two
equations 
\begin{equation*}
\sigma ^{\mu }a_{\mu }\psi =i\sigma ^{\mu }\partial _{\mu }\psi +m\psi \text{%
,\ }\sigma ^{\mu }a_{\mu }\psi =i\sigma ^{\mu }\partial _{\mu }\psi -m\psi .%
\text{\ \ }
\end{equation*}%
Substracting the second equation from the first one, and taking into acount
that $\psi \not\equiv 0$, we obtain $m=0$. Therefore both equations are
quivalent to the Weyl equation $\left( 2.2\right) $. In a similar way,
setting $\Psi =\left[ 
\begin{array}{c}
\psi \\ 
-\psi%
\end{array}%
\right] $ in $\left( 2.1\right) $ we get equation $\left( 2.3\right) $.
\end{proof}

Now we are ready \ to prove the main Theorem of this paper:

\begin{theorem}
Let $\Psi \not\equiv 0$ be a Dirac solution corresponding to a mass $m$\ and
a 4-potential $a_{\mu }$. Then $\Psi $ is degenerate, if and only if $\ \Psi
^{\dagger }\gamma \Psi \equiv 0$. Specifically we have:
\end{theorem}

1. If $\Psi ^{\dagger }\gamma \Psi \not\equiv 0$, then $\Psi $ corresponds
to one and only one real potential $a_{\mu }$ given by $\left( 2.4\right) $
or $\left( 2.5\right) .$

2. If $\Psi ^{\dagger }\gamma \Psi \equiv 0$, and $\Psi ^{T}\gamma ^{2}\Psi
\not\equiv 0$, then $\Psi $ corresponds to an infinite number of real
potentials $b_{\mu }$ of the form 
\begin{equation}
b_{\mu }=a_{\mu }+f\theta _{\mu },\;  \tag{5.7}
\end{equation}%
\ where $f$ is any real function of $\ $the variables $x_{\mu }$ , while the
real parameters $\theta _{\mu }$ are given by the formulae

\begin{equation}
\left( \theta _{0},\theta _{1},\theta _{2},\theta _{3}\right) =\left( 1,-%
\frac{\Psi ^{T}\gamma ^{0}\gamma ^{1}\gamma ^{2}\Psi }{\Psi ^{T}\gamma
^{2}\Psi },-\frac{\Psi ^{T}\gamma ^{0}\Psi }{\Psi ^{T}\gamma ^{2}\Psi },%
\frac{\Psi ^{T}\gamma ^{0}\gamma ^{2}\gamma ^{3}\Psi }{\Psi ^{T}\gamma
^{2}\Psi }\right)  \tag{5.8}
\end{equation}

3. If $\Psi ^{\dagger }\gamma \Psi \equiv 0$, and $\Psi ^{T}\gamma ^{2}\Psi
\equiv 0$, then either $\Psi =$ $\left[ 
\begin{array}{c}
\psi \\ 
\psi%
\end{array}%
\right] $ or $\Psi =$ $\left[ 
\begin{array}{c}
\psi \\ 
-\psi%
\end{array}%
\right] $, where $\psi $ is solution to the Weyl equation in the form (2.2)
or (2.3) respectively. In this case the set of all real 4-potentials $b_{\mu
}$ corresponding to $\Psi $ is given by $\left( 3.1\right) $ or $\left(
3.2\right) $ respectively.

\begin{proof}
1. See Remark 2.2.

2. First we will show that $\Psi $ corresponds to any real or complex
4-potential $b_{\mu }$ as given by $\left( 5.7\right) :$ Clearly, equation $%
\left( 2.2\right) $ can be written as 
\begin{equation}
(a_{\mu }+f\theta _{\mu })\gamma ^{\mu }\Psi =i\partial _{\mu }\gamma ^{\mu
}\Psi -m\Psi +f\theta _{\mu }\gamma ^{\mu }\Psi  \tag{5.9}
\end{equation}%
After some algebraic calculations it is easy to verfy that for any spinor $%
\Psi $ the following identity holds:%
\begin{equation}
\theta _{\mu }\gamma ^{\mu }\Psi \equiv 0.  \tag{5.10}
\end{equation}%
Now, from $\left( 5.9\right) $ and $\left( 5.10\right) $ we conclude that $%
\Psi $ corresponds to $b_{\mu }$.

Next we will try to extract from equation $\left( 2.1\right) $ all
4-potentials $b_{\mu }$ corresponding to $\Psi $ :

Since $\Psi $ corresponds also to $b_{\mu }$ we have \ that 
\begin{equation}
b_{\mu }\gamma ^{\mu }\Psi =i\gamma ^{\mu }\partial _{\mu }\Psi -m\Psi  
\tag{5.11}
\end{equation}%
Substracting $\left( 2.1\right) $ from $\left( 5.11\right) $ we obtain 
\begin{equation}
\gamma ^{\mu }\left( b_{\mu }-a_{\mu }\right) \Psi =0.  \tag{5.12}
\end{equation}%
Multiplying $\left( 5.12\right) $ from the left succesivelly by $\Psi
^{T}\gamma ^{1}\gamma ^{2}$, $\Psi ^{T}$, $\Psi ^{T}\gamma ^{2}\gamma ^{3}$
and using the fact that the matrices $\gamma ^{1}\gamma ^{2}\gamma ^{3}$, $%
\gamma ^{1},\gamma ^{3}$ are antisymmetric, \ we obtain the following set of
equations: 
\begin{align*}
\left( b_{1}-a_{1}\right) \Psi ^{T}\gamma ^{2}\Psi & = & & -\left(
b_{0}-a_{0}\right) \Psi ^{T}\gamma ^{0}\gamma ^{1}\gamma ^{2}\Psi \text{,} \\
\left( b_{2}-a_{2}\right) \Psi ^{T}\gamma ^{2}\Psi & = & & -\left(
b_{0}-a_{0}\right) \Psi ^{T}\gamma ^{0}\Psi \text{,} \\
\left( b_{3}-a_{3}\right) \Psi ^{T}\gamma ^{2}\Psi & = & & \left(
b_{0}-a_{0}\right) \Psi ^{T}\gamma ^{0}\gamma ^{2}\gamma ^{3}\Psi \text{.}
\end{align*}%
Setting $b_{0}=a_{0}+f,$ in the above equations we arrive to \ $\left(
5.7\right) .$ Finally, from Lemma 5.1\ follows that the functions $\theta
_{1}$,$~\theta _{2}$, $\theta _{3}$ are real. Consequently, all the 4
-potentials given by $\left( 5.7\right) $ are real.

3.This last part of Theorem 5.4 follows from Theorem 3.1 and Lemma 5.3.
\end{proof}

\begin{remark}
It is known \cite{Eliezer}, \cite{Booth-Legg-Jarvis} that any Dirac solution
corresponds to infinitelly many complex 4-potentials. Indeed, in the case
that $\Psi ^{\dagger }\gamma \Psi \not\equiv 0$, and $\Psi ^{T}\gamma
^{2}\Psi \not\equiv 0,$ according to Lemma 5.1, the functions $\theta _{1}$,$%
~\theta _{2}$, $\theta _{3}$ are complex and consequently the set of all
complex 4-potentials $b_{\mu }$ corresponding to $\Psi $ are given by $%
\left( 5.7\right) $ , where $f$ is an arbitrary complex function.
\end{remark}

\begin{remark}
\bigskip Let $\Psi $ be any degenerate Dirac solution such that $\Psi
^{\dagger }\gamma \Psi \equiv 0$, and $\Psi ^{T}\gamma ^{2}\Psi \not\equiv 0$%
. Then there are at least two different electromagnetic fields corresponding
to $\Psi $.
\end{remark}

\begin{proof}
We suppose that $\Psi $ corresponds to the 4-potential $a_{\mu }$. Then, if
we set $f=1$ and $f=x_{0}$, from Theorem (5.4) follows that $\Psi $ also
corresponds to the 4-potentials $b_{\mu }=\alpha _{\mu }+\theta _{\mu }$, $%
c_{\mu }=\alpha _{\mu }+x_{\mu }\theta _{\mu }$ respectively, where the
parameters $\theta _{\mu }$ are given by equation $\left( 5.8\right) $. We
also suppose that the 4-potentials $\alpha _{\mu },~b_{\mu },~c_{\mu }$ are
associated \ with a common electromagnetic field. Then the fields $\left(
1,\theta _{1},\theta _{2},\theta _{3}\right) $, $\left( x_{0},x_{0}\theta
_{1},x_{0}\theta _{2},x_{0}\theta _{3}\right) $ are conservative. Therefore,
we have that 
\begin{equation}
\partial _{0}\theta _{\mu }=0\text{,}\mu =1,2,3,\;  \tag{5.13}
\end{equation}%
and 
\begin{equation}
x_{0}\partial _{0}\theta _{\mu }+\theta _{\mu }=0,\mu =1,2,3.  \tag{5.14}
\end{equation}%
From $(5.13)$ and $\left( 5.14\right) $ follows that $\theta _{\mu }=0,$ for 
$\mu =1,2,3$. Then, using equations (5.8) it can be easily shown that the
spinors $\Psi $ must be of the form: 
\begin{equation*}
\Psi =\left[ 
\begin{array}{c}
\psi _{1} \\ 
\psi _{2} \\ 
\psi _{1} \\ 
\psi _{2}%
\end{array}%
\right] \text{ or }\Psi =\left[ 
\begin{array}{c}
\psi _{1} \\ 
\psi _{2} \\ 
-\psi _{1} \\ 
-\psi _{2}%
\end{array}%
\right] \text{.}
\end{equation*}%
which clearly satisfy: $\Psi ^{T}\gamma ^{2}\Psi \equiv 0$. This contradicts
the condition that $\Psi ^{T}\gamma ^{2}\dot{\Psi}\not\equiv 0$. Therefore
at least one of the two 4-potentials $b_{\mu }$ and $c_{\mu }$ is gauge
inequivalent to $a_{\mu }$.
\end{proof}

\begin{remark}
It is easy to prove, that $\Psi ^{\dagger }\gamma \Psi =0$ if and only if $%
\Psi $ has the following form, 
\begin{equation}
\Psi =u\left[ 
\begin{array}{c}
\overline{w} \\ 
1 \\ 
\overline{w} \\ 
1%
\end{array}%
\right] +v\left[ 
\begin{array}{c}
1 \\ 
-w \\ 
-1 \\ 
w%
\end{array}%
\right] \text{,}  \tag{5.15}
\end{equation}
\end{remark}

where $u,v,w$ are arbitrary functions of $\ x_{\mu }$.

Closing this section we should note that the property of Dirac particles,
described by spinors satisfying the condition $\Psi ^{\dagger }\gamma \Psi
=0,$ to be in the same state under a wide variety of different
electromagnetic fields is a particularly interesting and surprising result,
at least in the framework of classical Physics, where one would expect that
any changes to the electromagnetic fields, and consequently to the
electromagnetic forces acting on the particles, would alter their state.
However, as it will be shown in the following section massless Dirac
particles can exist in the same state both in the absence of electromagnetic
fields and in a wide variety of different electromagnetic fields, which are
explicitely calculated. The degeneracy can also be extended to massive
particles, when particle-antiparticle pairs are considered.

\section{D\textbf{egenerate solutions to the force-free Dirac equation and
the corresponding electromagnetic fields}}

In this section we focus on the degenerate solutions to the force-free Dirac
equation providing details regarding their structure as well as their
physical interpretation. We also provide explicit expressions regarding the
electromagnetic fields corresponding to these solutions, as well as the
electric charge and current densities required to produce these fields. This
analysis leads to the surprising result that a massless Dirac particle can
be in the same state either in free space (zero electromagnetic field) or in
a region of space with constant electric charge and current densities.

The general solution to the force-free Dirac equation for a particle of mass 
$m$, energy $E$, and momentum $\mathbf{p}=\left\vert \mathbf{p}\right\vert
\left( \sin \theta \cos \varphi \text{ }\mathbf{i}+\sin \theta \sin \varphi 
\text{ }\mathbf{j}+\cos \theta \text{ }\mathbf{k}\right) =p_{x}\mathbf{i}%
+p_{y}\mathbf{j}+p_{z}\mathbf{k}$\textbf{,} \ propagating along a direction
defined by the angles $\theta $,$\varphi $ in spherical coordinates can be
written in the form \cite{Thomson}

\begin{equation}
\Psi _{p}\left( \mathbf{r},t\right) =\left[ c_{1}u_{1}\left( E,\mathbf{p}%
\right) +c_{2}u_{2}\left( E,\mathbf{p}\right) \right] \exp \left[ i\left(
p_{x}x+p_{y}y+p_{z}z-Et\right) \right]  \tag{6.1}
\end{equation}

\bigskip where the 4-vectors $u_{1}\left( E,\mathbf{p}\right) $ , $%
u_{2}\left( E,\mathbf{p}\right) $ are eigenstates of the helicity operator
and are explicitly given by the formulae \bigskip 
\begin{equation}
u_{1}\left( E,\mathbf{p}\right) =\left( 
\begin{array}{c}
\cos \left( \frac{\theta }{2}\right) \\ 
e^{i\varphi }\sin \left( \frac{\theta }{2}\right) \\ 
\frac{\left\vert \mathbf{p}\right\vert }{E+m}\cos \left( \frac{\theta }{2}%
\right) \\ 
\frac{\left\vert \mathbf{p}\right\vert }{E+m}e^{i\varphi }\sin \left( \frac{%
\theta }{2}\right)%
\end{array}%
\right)  \tag{6.2}
\end{equation}

corresponding to a particle with positive helicity (spin parallel to its
propagation direction), and

\begin{equation}
u_{2}\left( E,\mathbf{p}\right) =\left( 
\begin{array}{c}
-\sin \left( \frac{\theta }{2}\right) \\ 
e^{i\varphi }\cos \left( \frac{\theta }{2}\right) \\ 
\frac{\left\vert \mathbf{p}\right\vert }{E+m}\sin \left( \frac{\theta }{2}%
\right) \\ 
-\frac{\left\vert \mathbf{p}\right\vert }{E+m}e^{i\varphi }\cos \left( \frac{%
\theta }{2}\right)%
\end{array}%
\right)  \tag{6.3}
\end{equation}

\bigskip corresponding to a particle with negative helicity (spin
anti-parallel to its propagation direction). Here $\mathbf{i,j,k}$ are the
unit vectors along the x-, y-, z-direction respectively in a Cartesian
coordinate system. The complex constants $c_{1},c_{2}$ correspond to the
contribution of the positive and negative helicity states respectively to
the actual state of the particle. Similarly, the general solution to the
force free Dirac equation for an antiparticle is \cite{Thomson}

\begin{equation}
\Psi _{a}\left( \mathbf{r},t\right) =\left[ c_{1}v_{1}\left( E,\mathbf{p}%
\right) +c_{2}v_{2}\left( E,\mathbf{p}\right) \right] \exp \left[ -i\left(
p_{x}x+p_{y}y+p_{z}z-Et\right) \right]  \tag{6.4}
\end{equation}

where

\begin{equation}
v_{1}\left( E,\mathbf{p}\right) =\left( 
\begin{array}{c}
\frac{\left\vert \mathbf{p}\right\vert }{E+m}\sin \left( \frac{\theta }{2}%
\right) \\ 
-\frac{\left\vert \mathbf{p}\right\vert }{E+m}e^{i\varphi }\cos \left( \frac{%
\theta }{2}\right) \\ 
-\sin \left( \frac{\theta }{2}\right) \\ 
e^{i\varphi }\cos \left( \frac{\theta }{2}\right)%
\end{array}%
\right)  \tag{6.5}
\end{equation}

corresponding to an anti-particle with positive helicity, and

\begin{equation}
v_{2}\left( E,\mathbf{p}\right) =\left( 
\begin{array}{c}
\frac{\left\vert \mathbf{p}\right\vert }{E+m}\cos \left( \frac{\theta }{2}%
\right) \\ 
\frac{\left\vert \mathbf{p}\right\vert }{E+m}e^{i\varphi }\sin \left( \frac{%
\theta }{2}\right) \\ 
\cos \left( \frac{\theta }{2}\right) \\ 
e^{i\varphi }\sin \left( \frac{\theta }{2}\right)%
\end{array}%
\right)  \tag{6.6}
\end{equation}

\bigskip corresponding to an anti-particle with negative helicity.\bigskip

In order these solutions to be degenerate we apply the condition $\Psi
^{\dagger }\gamma \Psi =0$ and find that it is valid if and only if \bigskip 
\begin{equation*}
\left\vert \mathbf{p}\right\vert ^{2}=\left( E+m\right) ^{2}\rightarrow
E^{2}-m^{2}=E^{2}+m^{2}+2Em\rightarrow m^{2}+Em=0\rightarrow \left\{ 
\begin{array}{c}
m=0 \\ 
E=-m%
\end{array}%
\right\}
\end{equation*}

\bigskip where we have also used the well-known formula of special
relativity $E^{2}=c^{2}\left\vert \mathbf{p}\right\vert ^{2}+m^{2}c^{4\text{ 
}}$which in the natural system of units $\left( \hbar =c=1\right) $ becomes $%
E^{2}=\left\vert \mathbf{p}\right\vert ^{2}+m^{2}{}^{\text{ }}.$ Obviously,
the condition $E=-m$, corresponding to particles or antiparticles with
negative mass, or negative energy, cannot be valid because in this case some
terms of the 4-vectors $u_{1},u_{2},v_{1},v_{2}$ become infinite. Therefore,
we are led to the very interesting result, that the wavefunction of a free
Dirac particle, or antiparticle, is degenerate if and only if the particle,
or antiparticle, is massless. At this point` it should be noted that, to our
knowledge, massless charged particles have not been discovered yet. However,
in the initial stages of the evolution of the Universe, before the particles
acquire mass, there was a plethora of massless charged particles \cite%
{Gromov}, \cite{Gorbunov}. Further, it has been shown recently \cite%
{Novoselov}, \cite{Ciudad}, \cite{Park}, \cite{Kamal}, \cite{Lai}, \cite%
{Armitage}, that in certain materials as graphene sheets, Weyl semimetals,
etc ordinary charged particles can also behave as massless. Therefore, our
analysis could provide new insight to the study of the initial phases of the
evolution of the universe as well as the behavior of "massless" particles in
new exotic materials.

In the case of massless particles, the general form of the degenerate
wavefunctions is

\begin{eqnarray}
\Psi _{p}\left( E,\mathbf{p}\right) &=&\left( 
\begin{array}{c}
c_{1}\cos \left( \frac{\theta }{2}\right) -c_{2}\sin \left( \frac{\theta }{2}%
\right) \\ 
e^{i\varphi }\left( c_{1}\sin \left( \frac{\theta }{2}\right) +c_{2}\cos
\left( \frac{\theta }{2}\right) \right) \\ 
c_{1}\cos \left( \frac{\theta }{2}\right) +c_{2}\sin \left( \frac{\theta }{2}%
\right) \\ 
e^{i\varphi }\left( c_{1}\sin \left( \frac{\theta }{2}\right) -c_{2}\cos
\left( \frac{\theta }{2}\right) \right)%
\end{array}%
\right)  \TCItag{6.7} \\
&&\times \exp \left[ iE\left( x\sin \theta \cos \varphi +y\sin \theta \sin
\varphi +z\cos \theta -t\right) \right]  \notag
\end{eqnarray}

\begin{eqnarray}
\Psi _{a}\left( E,\mathbf{p}\right) &=&\left( 
\begin{array}{c}
c_{1}\sin \left( \frac{\theta }{2}\right) +c_{2}\cos \left( \frac{\theta }{2}%
\right) \\ 
e^{i\varphi }\left( -c_{1}\cos \left( \frac{\theta }{2}\right) +c_{2}\sin
\left( \frac{\theta }{2}\right) \right) \\ 
-c_{1}\sin \left( \frac{\theta }{2}\right) +c_{2}\cos \left( \frac{\theta }{2%
}\right) \\ 
e^{i\varphi }\left( c_{1}\cos \left( \frac{\theta }{2}\right) +c_{2}\sin
\left( \frac{\theta }{2}\right) \right)%
\end{array}%
\right)  \TCItag{6.8} \\
&&\times \exp \left[ -iE\left( x\sin \theta \cos \varphi +y\sin \theta \sin
\varphi +z\cos \theta -t\right) \right]  \notag
\end{eqnarray}

\bigskip for particles and anti-particles respectively. These expressions
are easily obtained from equations (6.1)-(6.6) setting $m=0$ and taking into
account the fact that in this case $\left\vert \mathbf{p}\right\vert =E$ (in
the natural system of units where $\hbar =c=1$). The parameters $\theta
_{1},\theta _{2},\theta _{3}$, defined by Eq. (5.8) in the framework of
Theorem 5.4, corresponding to the above general degenerate solutions of the
force-free Dirac equation, are given by the particularly simple expressions

\begin{equation*}
\theta _{1}=-\sin \theta \cos \varphi ,\text{ }\theta _{2}=-\sin \theta \sin
\varphi ,\text{ }\theta _{3}=-\cos \theta
\end{equation*}

\bigskip which are obviously the opposite to the projections on the -x, -y
and -z axis respectively of a unit vector along the direction defined by the
angles $\left( \theta ,\varphi \right) $. Then, from Theorem 5.4, it is
concluded that the wavefunctions (6.7), (6.8) can describe particles and
anti-particles both in space free of electromagnetic fields, as well as in
space with non-zero electromagnetic fields related to the 4-potentials

\begin{equation*}
\left( f,-f\sin \theta \cos \varphi ,-f\sin \theta \sin \varphi ,-f\cos
\theta \right)
\end{equation*}

where $g\left( \mathbf{r},t\right) $ is an arbitrary function of the spatial
variables and time. Defining $\ g\left( \mathbf{r},t\right) =\frac{1}{q}%
f\left( \mathbf{r},t\right) $ \ the electric and magnetic potentials
corresponding to the above 4-potentials are\bigskip 
\begin{eqnarray*}
\varphi \left( \mathbf{r},t\right) &=&g\left( \mathbf{r},t\right) ,\text{ \ }
\\
\mathbf{A}\left( \mathbf{r},t\right) &=&g\left( \mathbf{r},t\right) \sin
\theta \cos \varphi \text{ }\mathbf{i}+g\left( \mathbf{r},t\right) \sin
\theta \sin \varphi \text{ }\mathbf{j}+g\left( \mathbf{r},t\right) \cos
\theta \text{ }\mathbf{k}
\end{eqnarray*}

\bigskip respectively. Thus, the electric and magnetic fields (in Gaussian
units) derived from the above 4-potential are explicitly given by the
formulae \cite{Jackson}

\begin{equation}
\mathbf{E}\left( \mathbf{r},t\right) =-\nabla \varphi -\frac{\partial 
\mathbf{A}}{\partial t}=-\nabla g-\frac{\partial g}{\partial t}\left( \sin
\theta \cos \varphi \text{ }\mathbf{i}+\sin \theta \sin \varphi \text{ }%
\mathbf{j}+\cos \theta \text{ }\mathbf{k}\right)  \tag{6.9}
\end{equation}

\begin{eqnarray}
\mathbf{B}\left( \mathbf{r},t\right) &=&\nabla \times \mathbf{A}=\left( \cos
\theta \frac{\partial g}{\partial y}-\sin \theta \sin \varphi \frac{\partial
g}{\partial z}\right) \mathbf{i}  \TCItag{6.10} \\
&&-\left( \cos \theta \frac{\partial g}{\partial x}-\sin \theta \cos \varphi 
\frac{\partial g}{\partial z}\right) \mathbf{j}+\sin \theta \left( \sin
\varphi \frac{\partial g}{\partial x}-\cos \varphi \frac{\partial g}{%
\partial y}\right) \mathbf{k}  \notag
\end{eqnarray}

\bigskip where we have set $c=1,$ assuming that we are working in the
natural system of units.

Thus, a massless Dirac particle propagating along the direction defined by
the angles $\left( \theta ,\varphi \right) $ will be in the same state both
in free space and in any electromagnetic field determined by equations
(6.9), (6.10). At this point we can proceed a step further calculating the
electric charge $\rho \left( \mathbf{r},t\right) $ and the current\ $\mathbf{%
J}\left( \mathbf{r},t\right) $ densities required to produce the above
electromagnetic fields, using Maxwell's equations in the form \cite{Jackson}

\begin{equation*}
\nabla ^{2}\varphi +\frac{\partial }{\partial t}\left( \nabla \cdot \mathbf{A%
}\right) =-4\pi \rho
\end{equation*}

\begin{equation*}
\left( \nabla ^{2}\mathbf{A}-\frac{\partial ^{2}\mathbf{A}}{\partial t^{2}}%
\right) -\nabla \left( \nabla \cdot \mathbf{A}+\frac{\partial \varphi }{%
\partial t}\right) =-4\pi \mathbf{J}
\end{equation*}

\bigskip where we have also set $c=1,$ assuming that we are working in the
natural system of units.

Then, it is easy to show that the electric charge and current densities
required to produce the electromagnetic fields (6.9), (6.10) are given, in
cartesian coordinates, by the formulae

\begin{eqnarray}
\rho \left( \mathbf{r},t\right) &=&-\frac{1}{4\pi }\nabla ^{2}g 
\TCItag{6.11} \\
&&-\frac{1}{4\pi }\frac{\partial }{\partial t}\left( \sin \theta \cos
\varphi \frac{\partial g}{\partial x}+\sin \theta \sin \varphi \frac{%
\partial g}{\partial y}+\cos \theta \frac{\partial g}{\partial z}\right) 
\notag
\end{eqnarray}

and

\begin{eqnarray}
J_{x}\left( \mathbf{r},t\right) &=&-\frac{1}{4\pi }\sin \theta \cos \varphi
\left( \nabla ^{2}g-\frac{\partial ^{2}g}{\partial t^{2}}\right) 
\TCItag{6.12} \\
&&+\frac{1}{4\pi }\frac{\partial }{\partial x}\left( \sin \theta \cos
\varphi \frac{\partial g}{\partial x}+\sin \theta \sin \varphi \frac{%
\partial g}{\partial y}+\cos \theta \frac{\partial g}{\partial z}+\frac{%
\partial g}{\partial t}\right)  \notag
\end{eqnarray}

\begin{eqnarray}
J_{y}\left( \mathbf{r},t\right) &=&-\frac{1}{4\pi }\sin \theta \sin \varphi
\left( \nabla ^{2}g-\frac{\partial ^{2}g}{\partial t^{2}}\right) 
\TCItag{6.13} \\
&&+\frac{1}{4\pi }\frac{\partial }{\partial y}\left( \sin \theta \cos
\varphi \frac{\partial g}{\partial x}+\sin \theta \sin \varphi \frac{%
\partial g}{\partial y}+\cos \theta \frac{\partial g}{\partial z}+\frac{%
\partial g}{\partial t}\right)  \notag
\end{eqnarray}

\begin{eqnarray}
J_{z}\left( \mathbf{r},t\right) &=&-\frac{1}{4\pi }\cos \theta \left( \nabla
^{2}g-\frac{\partial ^{2}g}{\partial t^{2}}\right)  \TCItag{6.14} \\
&&+\frac{1}{4\pi }\frac{\partial }{\partial z}\left( \sin \theta \cos
\varphi \frac{\partial g}{\partial x}+\sin \theta \sin \varphi \frac{%
\partial g}{\partial y}+\cos \theta \frac{\partial g}{\partial z}+\frac{%
\partial g}{\partial t}\right)  \notag
\end{eqnarray}

\bigskip Thus, a massless Dirac particle propagating along the direction
defined by the angles $\left( \theta ,\varphi \right) $ in spherical
coordinates will be in the same state both in free space and in any
electromagnetic field determined by equations (6.9), (6.10), which is
produced by the electric charge and current densities given by equations
(6.11)-(6.14). We should note that, as it is easily deduced from the above
equations, if the arbitrary function $g$ is time independent, then the
electric charge and current densities become also time-independent,
corresponding to a static electromagnetic field.

A special case of particular interest is when $g\left( \mathbf{r}\right)
=c_{0}\left\vert \mathbf{r}\right\vert ^{2}=c_{0}\left(
x^{2}+y^{2}+z^{2}\right) $, where $c_{0}$ is an arbitrary real constant.
Then, it is easy to show that the electric charge and current densities take
the particularly simple form

\begin{equation*}
\rho =-\frac{3c_{0}}{2\pi },\text{ \ }\mathbf{J}=-\frac{c_{0}}{\pi }\left(
\sin \theta \cos \varphi \text{ }\mathbf{i}+\sin \theta \sin \varphi \text{ }%
\mathbf{j}+\cos \theta \text{ }\mathbf{k}\right)
\end{equation*}

\bigskip leading to the surprising result that a massless charged particle
propagating along the direction defined by the angles $\left( \theta
,\varphi \right) $ will be in the same state either in free space (zero
electromagnetic field) or in a region of space with constant electric charge
and current densities given by the above formulae.

Finally, it should be noted that in the case of particle-antiparticle pairs,
one can obtain degenerate solutions even for massive particles. Indeed, it
is easy to show that if $\left\vert c_{1}\right\vert =\left\vert
c_{2}\right\vert ,$ the spinors

\begin{eqnarray*}
\Psi _{\uparrow }\left( \mathbf{r},t\right) &=&c_{1}u_{1}\left( E,\mathbf{p}%
\right) \exp \left[ i\left( p_{x}x+p_{y}y+p_{z}z-Et\right) \right] \\
&&+c_{2}v_{1}\left( E,\mathbf{p}\right) \exp \left[ -i\left(
p_{x}x+p_{y}y+p_{z}z-Et\right) \right]
\end{eqnarray*}

\begin{eqnarray*}
\Psi _{\downarrow }\left( \mathbf{r},t\right) &=&c_{1}u_{2}\left( E,\mathbf{p%
}\right) \exp \left[ i\left( p_{x}x+p_{y}y+p_{z}z-Et\right) \right] \\
&&+c_{2}v_{2}\left( E,\mathbf{p}\right) \exp \left[ -i\left(
p_{x}x+p_{y}y+p_{z}z-Et\right) \right]
\end{eqnarray*}

corresponding to particle-antiparticle pairs with positive and negative
helicity respectively, are degenerate, for any value of the mass $m.$ Here, $%
c_{1},c_{2}$ are arbitrary complex constants satisfying the condition $%
\left\vert c_{1}\right\vert =\left\vert c_{2}\right\vert ,$ while $%
u_{1},u_{2},v_{1},v_{2}$ are defined by (6.2), (6.3), (6.5), (6.6)
respectively. More details on these solutions will be provided in a future
work.

\section{\textbf{\ Solutions to the Weyl equation and the corresponding
electromagnetic fields}}

\bigskip In this section we shall provide explicit expressions regarding the
infinite number of electromagnetic fields corresponding to specific classes
of Weyl spinors. First we consider the force-free Weyl equation in the form $%
(2.2)$ where all the components of the electromagnetic 4-potential are set
equal to zero. In this case it is easy to show that the spinor%
\begin{equation}
\Psi =\left( 
\begin{array}{c}
\cos \left( \frac{\theta }{2}\right) \\ 
e^{i\varphi }\sin \left( \frac{\theta }{2}\right)%
\end{array}%
\right) \exp \left[ iE\left( x\sin \theta \cos \varphi +y\sin \theta \sin
\varphi +z\cos \theta -t\right) \right]  \tag{7.1}
\end{equation}

\bigskip is solution to the force-free Weyl equation corresponding to a free
Weyl particle of energy $E$, propagating along a direction defined by the
angles $\left( \theta ,\varphi \right) $ in spherical coordinates. Then,
according to Theorem 3.1, the spinor (7.1) is also solution to the Weyl
equation for the following set of 4-potentials\bigskip 
\begin{equation*}
\left( f,\varphi _{1}f,\varphi _{2}f,\varphi _{3}f\right)
\end{equation*}

\bigskip where $f\left( \mathbf{r},t\right) $ is an arbitrary function of
the spatial coordinates and time. The parameters $\varphi _{1},\varphi
_{2},\varphi _{3}$ can be easily calculated taking the values

\begin{equation*}
\varphi _{1}=-\sin \theta \cos \varphi ,\text{ }\varphi _{2}=-\sin \theta
\sin \varphi ,\text{ }\varphi _{3}=-\cos \theta
\end{equation*}

\bigskip which are obviously the opposites to the projections of a unit
vector along the direction defined by the angles $\left( \theta ,\varphi
\right) $ on the -x, -y and -z axis respectively. Defining the function $\
g\left( \mathbf{r},t\right) =\frac{1}{q}f\left( \mathbf{r},t\right) $ \ the
electric and magnetic potentials corresponding to the above 4-potentials
are\bigskip 
\begin{eqnarray*}
\varphi \left( \mathbf{r},t\right) &=&g\left( \mathbf{r},t\right) ,\text{ \ }
\\
\mathbf{A}\left( \mathbf{r},t\right) &=&g\left( \mathbf{r},t\right) \sin
\theta \cos \varphi \text{ }\mathbf{i}+g\left( \mathbf{r},t\right) \sin
\theta \sin \varphi \text{ }\mathbf{j}+g\left( \mathbf{r},t\right) \cos
\theta \text{ }\mathbf{k}
\end{eqnarray*}

\bigskip respectively. Here it is assumed that we are working \ in the
natural system of units where $\hbar =c=1$. The electric and magnetic fields
(in Gaussian units) derived from the above 4-potential are explicitly given
by the formulae \cite{Jackson}

\begin{equation}
\mathbf{E}\left( \mathbf{r},t\right) =-\nabla \varphi -\frac{\partial 
\mathbf{A}}{\partial t}=-\nabla g-\frac{\partial g}{\partial t}\left( \sin
\theta \cos \varphi \text{ }\mathbf{i}+\sin \theta \sin \varphi \text{ }%
\mathbf{j}+\cos \theta \text{ }\mathbf{k}\right)  \tag{7.2}
\end{equation}

\begin{eqnarray}
\mathbf{B}\left( \mathbf{r},t\right) &=&\nabla \times \mathbf{A}=\left( \cos
\theta \frac{\partial g}{\partial y}-\sin \theta \sin \varphi \frac{\partial
g}{\partial z}\right) \mathbf{i}  \TCItag{7.3} \\
&&-\left( \cos \theta \frac{\partial g}{\partial x}-\sin \theta \cos \varphi 
\frac{\partial g}{\partial z}\right) \mathbf{j}+\sin \theta \left( \sin
\varphi \frac{\partial g}{\partial x}-\cos \varphi \frac{\partial g}{%
\partial y}\right) \mathbf{k}  \notag
\end{eqnarray}

Thus, a free Weyl particle propagating along the direction defined by the
angles $\left( \theta ,\varphi \right) $ will be in the same state both in
free space (zero electromagnetic field) as well as in any electromagnetic
field of the form described by equations (7.2), (7.3). Setting $\bigskip
g\left( \mathbf{r},t\right) =s\left( t\right) $ in the above fomulae, where $%
s\left( t\right) $ is an arbitrary real function of time, the magnetic field
becomes zero, while the electric field takes the simple form

\begin{equation*}
\mathbf{E}\left( \mathbf{r},t\right) =-\frac{ds}{dt}\left( \sin \theta \cos
\varphi \text{ }\mathbf{i}+\sin \theta \sin \varphi \text{ }\mathbf{j}+\cos
\theta \text{ }\mathbf{k}\right)
\end{equation*}%
Obviously, if $s\left( t\right) =c_{0}t,$ where $c_{0}$ is an arbitrary real
constant, the above formula corresponds to a constant electric field.
Consequently, a constant or time - dependent electric field parallel (or
antiparallel) to the direction of motion of a Weyl particle does not alter
its state, and it will keep on moving with the same momentum as if there was
no electric field. Thus, Ohm's law does not hold for Weyl paticles, and the
current transferred by them \ remains constant, independent of the applied
electric field. This practically means that the resistance of a Weyl
material "adjusts" to the applied electric field in a way that the current
remains constant. This non-conventional behavior of Weyl materials regarding
their interaction with electric fields is expected to offer new
opportunities in nanoelectronics. As it can be deduced from (6.9), (6.10),
this remarkable result is also true for massless Dirac particles. Finally,
it should be mentioned that the property of massless particles to maintain
their state under a constant or time-dependent electric field, is closely
related to the fact that they move at a constant speed, either the speed of
light if moving in vacuum, or a much smaller value if moving in materials 
\cite{Lai}.

In the following we shall study the case of a Weyl particle confined in one
dimension by a constant magnetic field. Specifically, it is easy to show
that any spinor of the form

\begin{equation}
\Psi =h\left( y\right) \left( 
\begin{array}{c}
0 \\ 
1%
\end{array}%
\right) d\left( z+t\right)  \tag{7.4}
\end{equation}

where $h\left( y\right) $ is an arbitrary real function of $y$ and $d\left(
z+t\right) $ an arbitrary complex function of $z+t$ , is solution to the
Weyl equation for the 4-potential\bigskip

\begin{equation*}
\left( 0,-\frac{h_{y}}{h},0,0\right)
\end{equation*}

\bigskip where $h_{y}=\frac{dh\left( y\right) }{dy}.$Defining $\ H\left(
y\right) =-\frac{1}{q}\frac{h_{y}}{h}$\ it is straightforward to show that
the electromagnetic field corresponding to the above 4-potential is\qquad
\qquad \qquad \qquad \qquad\ 
\begin{equation*}
\mathbf{E}=0,\text{ \ }\mathbf{B}=\frac{dH}{dy}\mathbf{k}=-\frac{1}{q}\frac{%
h_{yy}h-h_{y}^{2}}{h^{2}}\mathbf{k}
\end{equation*}

\bigskip According to Theorem 3.1, the spinor (7.4) is also solution to the
Weyl equation for the 4-potentials

\begin{equation*}
\left( f,-\frac{h_{y}}{h},0,f\right)
\end{equation*}

\bigskip where $f\left( \mathbf{r},t\right) $ is an arbitrary real function
of the spatial coordinates and time. Defining $\ g\left( \mathbf{r},t\right)
=\frac{1}{q}f\left( \mathbf{r},t\right) $ it is straightforward to show that
a Weyl particle described by the spinor (7.4) will be in the same state
under the unfluence of any of the following electromagnetic fields

\bigskip 
\begin{eqnarray}
\mathbf{E}\left( \mathbf{r},t\right) &=&-\frac{\partial g}{\partial x}%
\mathbf{i}-\frac{\partial g}{\partial y}\mathbf{j}-\left( \frac{\partial g}{%
\partial z}-\frac{\partial g}{\partial t}\right) \mathbf{k}\text{ } 
\TCItag{7.5} \\
\text{\ }\mathbf{B}\left( \mathbf{r},t\right) &=&-\frac{\partial g}{\partial
y}\mathbf{i}+\frac{\partial g}{\partial x}\mathbf{j}+\frac{\partial H}{%
\partial y}\mathbf{k}  \TCItag{7.6}
\end{eqnarray}

As a specific example we consider the case where $h\left( y\right) $ is a
Gaussian function of the form$\ h\left( y\right) =\exp \left( -\lambda
y^{2}\right) $, where $\lambda $ is an arbitrary positive real constant, and 
$d\left( z+t\right) $ is a wave function of the form $d\left( z+t\right)
=A\exp \left[ -iE\left( z+t\right) \right] $, where $A$ is an arbitrary
complex constant and $E$ a positive real constant corresponding to the
energy of the particle. In this case the spinor \bigskip 
\begin{equation*}
\Psi =A\left( 
\begin{array}{c}
0 \\ 
1%
\end{array}%
\right) \exp \left[ -\lambda y^{2}-iE\left( z+t\right) \right]
\end{equation*}

\bigskip corresponds to a Weyl particle of energy $E$ propagating along the $%
-z$ direction, which is confined along the y-axis by a constant magnetic
field, given by the formula

\begin{equation*}
\mathbf{B}=\frac{dH}{dy}\mathbf{k}=\frac{2\lambda }{q}\mathbf{k}
\end{equation*}

Then, according to equations (7.5), (7.6) this particle will be in the same
state under the influence of an infinite number of electromagnetic fields of
the form

\bigskip 
\begin{eqnarray*}
\mathbf{E}\left( \mathbf{r},t\right) &=&-\frac{\partial g}{\partial x}%
\mathbf{i}-\frac{\partial g}{\partial y}\mathbf{j}-\left( \frac{\partial g}{%
\partial z}+\frac{\partial g}{\partial t}\right) \mathbf{k}\text{ } \\
\text{\ }\mathbf{B}\left( \mathbf{r},t\right) &=&\frac{\partial g}{\partial y%
}\mathbf{i}-\frac{\partial g}{\partial x}\mathbf{j}-\frac{2\lambda }{q}%
\mathbf{k}
\end{eqnarray*}

\bigskip where $g\left( \mathbf{r},t\right) =\frac{1}{q}f\left( \mathbf{r}%
,t\right) $ is an arbitrary real function of the spatial variables and time.
\ Finally, it should be mentioned that if the arbitrary function $g$ depends
only on time, the electric and magnetic fields become

\begin{equation*}
\mathbf{E}\left( \mathbf{r},t\right) =-\frac{dg}{dt}\mathbf{k},\text{ \ \ }%
\mathbf{B}\left( \mathbf{r},t\right) =\frac{2\lambda }{q}\mathbf{k}
\end{equation*}%
Thus, the state of the Weyl particle does not change if a constant or
time-dependent electric field is applied parallel to the magnetic field.
Similar results can be obtained regarding the alternative form of the Weyl
equation (2.3) for negative helicity particles.

From the examples discussed in sections 6, 7 it is clear that our theory has
very important physical implications as well as high potential for practical
applications. Specifically, we have shown that Weyl particles - as well as
massless Dirac particles - can be in the same state under a wide variety of
different electromagnetic fields. This result could be utilized to explain
the rapid evolution of the Universe during the first stages of its creation,
where all particles were massless \cite{Gromov}, \cite{Gorbunov}. Further,
the fact that massless Dirac and Weyl particles are very resistant to
electromagnetic perturbations is expected to play an important role
regarding the practical applications of new exotic materials, as graphene
sheets, Weyl semimetals, etc., where ordinary charged particles can behave
as massless \cite{Novoselov}, \cite{Ciudad}, \cite{Park}, \cite{Kamal}, \cite%
{Lai}, \cite{Armitage}. For example, as discussed above, one very important
property of these particles is that they remain in the same state under the
influence of a constant or time - dependent electric field parallel (or
antiparallel) to their direction of motion. This practically means that a
material where the current carriers are massless Dirac or Weyl particles
does not obey Ohm's law, in the sense that the resistance of this material
"adjusts" to the applied electric field in a way that the current remains
constant, independent of the applied voltage. Obviously, this
non-conventional behavior could be utilized in nanoelectronics for designing
novel components and devices.

\section{\textbf{Summary}}

\textbf{\ }\ \ In the present study we have first shown that all solutions
to the Weyl equations are degenerate, in the sense that they correspond to
an infinite number of electromagnetic 4-potentials, which are explicitely
provided. Further, it is thoroughly proven that every Dirac spinor
corresponds to one and only one mass.Using this result we have proven the
main theorem of this article, according to which all Dirac spinors can be
classified into two classes. The elements of the first class correspond to
one and only one 4-potential, and are called non-degenerate Dirac solutions,
while the elements of the second class correspond to an infinite number of
4-potentials, and are called degenerate Dirac solutions. Further, it has
been shown that at least two of these 4-potentials are gauge-inequivalent,
corresponding to different electromagnetic fields. In order to illustrate
this particularly important result we have studied the force-free Dirac
equation and found that the degenerate solutions correspond to massless
particles. Thus, a massless Dirac particle can be in the same state both in
the absence of electromagnetic fields as well as under the influence of an
infinite number of different electromagnetic fields, which are calculated.
It has also been shown that the degeneracy can be extended to massive
particles, if particle-antiparticle pairs are considered. Specific examples
of Weyl spinors are also provided, calculating the infinite number of
different electromagnetic fields corresponding to these solutions.This
property of massless particles to be in the same state under the influence
of different electromagnetic fields could be used to to explain the rapid
evolution of the Universe during the firts stages of its creation - before
the particles acquire mass - as well as the behavior of ordinary charged
particles in new exotic materials, e.g. graphene sheets, Weyl semimetals,
etc., where they can collectively behave as massless. Finally, we have shown
that Ohm's law does not hold for massless Dirac and Weyl particles, and the
current transferred by them is constant, independent of the applied electric
field.

\textbf{Acknowledgement \ }The authors wish to express their gratitude to
Prof David Fairlie for his insightful suggestions relevant to Prof Christie
Eliezer's earlier paper.

\end{document}